\documentclass{jaa}

\usepackage{graphicx,graphics,amsmath}
\usepackage{xcolor}
\usepackage[normalem]{ulem}

\usepackage{amsmath}	
\usepackage{amssymb}
\usepackage{amssymb,ulem}
\usepackage[T1]{fontenc}
\usepackage{ae,aecompl}
 \usepackage[sort&compress,round,comma,authoryear]{natbib}
\def\apj{{Astroph.\@ J.\ }}

\def\mnras{{Mon.\@ Not.\@ Roy.\@ Ast.\@ Soc. }}

\def\aj{{Astron.\@ J.\ }}

\def\prd{{Phys.\@ Rev.\@ D\ }}

\def\apjl{{Astroph.\@ J.\@ Lett.}}
\def\jcap{{JCAP}}
\def\apss{ApSS}
\def\apjs{ApJS}
\def\aap{AAp}

\begin{document}

\title{Exact Solutions and Constraints on the Dark Energy Model in FRW Universe }


\author{Albin Joseph\textsuperscript{1,*}, Rajib Saha\textsuperscript{1} }
\affilOne{\textsuperscript{1}Department of Physics, Indian Institute of Science Education and Research Bhopal, Bhopal 462 066, India\\}


\twocolumn[{

\maketitle

\corres{rajib@iiserb.ac.in}


\begin{abstract}
The inflationary epoch and the late time acceleration of the expansion rate of universe can be 
explained by assuming a gravitationally coupled scalar field. In this article, we propose 
a new method of finding exact solutions  in the background of flat Friedmann-Robertson-Walker (FRW) 
cosmological models by considering both scalar field and matter where the 
scalar field potential is a function of the scale factor. Our method provides analytical expressions 
for equation of state parameter of scalar field, deceleration parameter and Hubble parameter.
This method can be applied to various other forms of scalar field potential, to the early radiation 
dominated epoch and very early scalar field dominated inflationary dynamics. Since the method produces 
exact analytical expression for $H(a)$ (i.e., H(z) as well),  we then constrain the model with currents data sets, which includes-Baryon Acoustic Oscillations,
Hubble parameter data and Type 1a Supernova data (Pantheon Dataset). As an extension of the method, we also consider the inverse
problem of reconstructing scalar field potential energy by assuming
any general analytical expression of  scalar field equation
of state parameter as a function of scale factor. 
\end{abstract}

\keywords{Cosmology---Dark Energy---Exact Solutions.}

}]


\doinum{\#\#\#\#}
\artcitid{\#\#\#\#}
\volnum{\#\#\#\#}
\year{\#\#\#\#}

\section{Introduction}

The observations of type I-A supernovae indicate  that  expansion rate of the universe 
in the recent past (on cosmological time scale) is speeding up~\citep{1998AJ....116.1009R,1999ApJ...517..565P, 
	2004ApJ...606..702T,2007ApJS..170..377S,2007ApJ...666..716D,2008ApJ...686..749K,2009ApJ...700.1097H,2009ApJS..180..330K,
	2009ApJS..180..225H,2000MNRAS.317..893L,2009ApJ...690L..85L,2010ApJ...714L.185B,2011ApJS..192...18K,2014A&A...571A..16P,2020A&A...641A...6P}. 
This discovery serves  as a  paradigm shift  in our understanding of cosmology by postulating the 
existence of a component named `dark energy'. The analysis of current cosmological observations~\citep{2020A&A...641A...6P} 
indicates that the `dark energy' provides  dominant contribution to the present total energy density of the universe.
The accelerated expansion took place also in a widely separated time epoch, before the Universe became 
radiation dominated, during inflation~\citep{1981PhRvD..23..347G}, the theory of which was subsequently developed  by 
~\citep{1982PhLB..108..389L,1983PhLB..129..177L,1981PhRvD..23..347G}. The inflationary epoch as well as the recent accelerated 
expansion can be model led by postulating existence of a scalar field dynamically coupled to gravitation.
Since a scalar field is a simple, yet natural candidate which causes accelerated expansion it plays a fundamental role 
in cosmology. Scalar fields have been extensively studied in cosmology 
(see~\citep{1988PhRvD..37.3406R,1982PhLB..108..389L,1983PhLB..129..177L,2003RvMP...75..559P,2012Ap&SS.342..155B} and references therein). 
The scalar field, in this context, serves as the model of dark-energy. 
For the purpose of constraining the nature of dark-energy understanding the evolution of the universe
during the accelerated expansion epoch of universe is an area of great research interest.  

Currently, there is no unique underlying principle can uniquely specify the potential of the scalar 
field that gives rise to earlier inflationary epoch and the late time accelerated epoch of the 
universe. Many proposals based on new particle physics and gravitational theories were introduced 
(see \citep{2005hep.th....3203L} and references therein) and others were based on ad-hoc assumption so as 
to get the desired evolution of the universe \citep{1991CQGra...8..667E}. There is also a formalism where 
we can reconstruct the potential by using the knowledge of tensor gravitational spectrum and the 
scalar density fluctuation spectrum \citep{1993PhRvD..48.2529C,1994PhRvD..50..758L}. Though there are numerous 
scalar field potential which can give rise to accelerated expansion, the exact solutions of these 
cosmological models are less known. As the exact solutions of cosmological models gives rise to exact 
cosmological parameters, they have a vital role in the present cosmological scenario. There are several 
methods by which one can explore the exact solutions of Friedman equations in a scalar field dominated 
universe. The construction of exact solutions for an inflationary scenario was started with 
Muslimov~\citep{1990CQGra...7..231M}. The author, by starting with the assumption of scalar field potential $V(\phi)$, 
found the remaining parameters, i.e $a(t)$ and $\phi(t)$ based upon the model. A method of 
generating exact solutions in scalar field dominated cosmology by considering scalar field potential 
as a function of time $V(t)$ has been explored in \citep{1998JETP...87..223Z}. By assigning the time dependence 
of scale factor $a(t)$, we can also find the scalar field $\phi(t)$ and potential $V(t)$  as seen in \citep{1991CQGra...8..667E}. 
One can also find the analytical expressions for $a(t)$ and $V(\phi)$ by assigning time dependence of 
scalar field $\phi(t)$ which is explored in \citep{1993PhRvD..48.1585B}. Barrow \citep{1990PhLB..235...40B} showed a 
simple method of finding exact solutions of cosmological dynamic equations in terms of a pressure-density 
relationship.

One can also reduce the scalar field cosmology equations to a known type of equation whose solution 
has already been developed. In~\citep{2014EPJC...74.2784H,2017GReGr..49...24C} we can see a method in which the 
Klein-Gordon equation which describes the dynamics of the scalar field is transformed to a first order 
non-linear differential equation. This equation  immediately leads to the identification of some 
exact classes of scalar field potentials $V(\phi)$ for which the field equations can be solved exactly 
and there by obtaining analytical expressions for $a(t)$, $\phi(t)$ and $q(t)$ . The solutions of the  
Friedman equations in a scalar field dominated universe is explored by its connection with the Abel 
equations of first kind  is seen in \citep{2010JMP....51h2503Y}. Here for a given $V(\phi)$ one can obtain 
$a(t)$ and $\phi(t)$ analytically. The exact solutions for exponential form of the potential $V(\phi)$ 
by rewriting the Klein-Gordon equation in the Riccati form and thereby transforming it into a second-order 
linear differential equation is investigated in \citep{2011JCAP...10..004A}. Analytical solutions to the field 
equations can also be obtained  by considering suitable generating functions. Here the generating functions 
are chosen as a function of one of the parameters of the model \citep{2000PhRvD..61h7303K,2010JCAP...08..022C,2014EPJC...74.2784H,2018EPJC...78..301C} 
and thus by simplifying the scalar field cosmology equation one can obtain all the parameters of the model.  
In \citep{1990PhRvD..42.3936S} a method was proposed by simplifying scalar field cosmology equation by assuming 
Hubble function as a function of scalar field $\phi$. By making use of the Noether symmetry for exponential 
potential \citep{2014PhRvD..90j3524P,1990PhRvD..42.1091D}, Hojman's conservation law \cite{2013PhLB..726..471C}, 
and other non-canonical conservation laws \citep{2016PhRvD..93l3518D} for arbitrary potential and also by the 
form-invariant transformations of scalar field cosmology equations \citep{2013MPLA...2850235C} one can obtain the 
exact solutions for the parameters of the model.  Moreover analytical solutions for field equations by considering a homogeneous scalar field in the Szekeres cosmological metric has been investigated in~\citep{2018EPJC...78..767B}. 

Though there are various ways for finding exact solutions in scalar field cosmology, the solutions are limited 
if we incorporate the contributions by a perfect fluid source. In~\citep{1996IJMPD...5...71C} the authors showed that the Einstein's 
equations with a self-interacting minimally coupled scalar field, a perfect fluid source and cosmological constant 
can be reduced to quadrature in the Robertson-Walker metric. Here the scale factor is considered as the independent 
variable and the scalar field potential is expressed in terms of scale factor. Barrow \citep{1993CQGra..10..279B} presented 
exact solutions of Friedman universes which contain a scalar field and a perfect fluid with the requirement that 
the kinetic and potential energies of the scalar field be proportional. The classes of scalar field potentials 
$V(\phi)$, which provide exact solutions for scalar field with scaling behavior had been investigated 
in \citep{1999PhRvD..59b3509L}. In~\citep{2016PhRvD..94h3518B} specific closed-form solutions of field equations(with and without matter source) have been derived by assuming special inflationary functions for the scale factor and special equation of state parameters of the scalar field. The application of lie symmetry methods in finding the exact cosmological solutions for scalar field dark energy in the presence of perfect fluids has been investigated in~\citep{2015PhRvD..91l3535P}. In ~\citep{2015Ap&SS.360...20S}, a scenario of time varying cosmological term is investigated by using a special anartz were energy density of the scalar field is proportional to energy density of the barotropic fluid. Fomin \citep{2018RuPhJ..61..843F} explored the exact solutions by considering scalar field and matter 
fields or non-zero curvature by representing the main cosmological parameters as a function of number of $e$-folds 
and also by the direct substitution of the scale factor. The study of late-time cosmology in a (phantom) scalar-tensor theory with an exponential potential had been investigated in~\citep{2004PhRvD..70d3539E}. In \citep{2008PhRvD..77j6005E}, the unification of inflation and late-time acceleration epochs within the context of a single field theory had been studied. Moreover, the exact and semiclassical solutions of the Wheeler-DeWitt equation for a particular family of scalar field potential had been explored in \citep{2007IJMPD..16..641G}.

In this paper we propose a new method of finding analytical solutions of equation of 
state parameter of scalar field, deceleration parameter and also the Hubble parameter 
as a function of scale factor. Using the continuity equation for the scalar field we 
form a first order linear inhomogeneous equation of the independent variable  $\dot \phi^2$ 
and dependent variable  $a$. Since the inhomogeneous term is given by $-dV/da$, derivative of the 
scalar  field $a$ the linear equation is exactly solvable for any choice of $V = V(a)$. As 
some test applications we present solutions for some chosen forms of $V(a)$. Since the 
linear equation is completely decoupled from the other source terms of the Friedmann equation 
the solution of $\dot \phi^2(a)$ and  the chosen form of $V(a)$  can be used in the right 
hand side of first Friedmann equation together with the other source terms. Therefore,  apart 
from the scalar field, our  method can  incorporate matter or radiation as a perfect fluid source
to obtain an exact solution of $H(a)$ or $H(z)$, which can be constrained using observations. Since the method is applicable for any well behaved 
(e.g., differentiable  w.r.t. $a$) chosen form of $V(a)$ it can be applied  to understand the physics 
of the inflation, as well as the late time acceleration of expansion rate. 

The paper is organized as follows.  In Sec. \ref{sec:basic} we explore the basis equations 
in scalar field cosmology. In Sec. \ref{sec:power_law} we consider the scalar field potential 
energy of power law form. The Sec. \ref{sec:reconstruct} is dedicated to the reconstruction of scalar field potential energy. After discussing the relationship of the scalar field potential with  the particle physics models in Sec. \ref{sec:physical}, we constrain the model parameters in Sec. \ref{sec:constraints} Finally, 
in the last Sec. \ref{sec:conclusion} we discuss and conclude upon our results.

\section{Formalism}
\label{sec:basic}

In the spatially flat Friedman-Robertson-Walker (FRW) model of the universe the 
space-time interval $ds$ between two events in a global comoving Cartesian coordinate 
system follows, 
\begin{equation}
	ds^2=dt^2-a^2(t)\bigl(dx^2+dy^2+dz^2\bigr) \, , 
\end{equation}
where $a$  and $t$ represent  the scale factor and comoving time respectively and we have used
$c =1$ unit.  If the universe is dominated by the non-relativistic matter (e. g., baryon and cold dark matter) 
and a spatially homogeneous and time varying scalar field, $\phi$ which is minimally coupled to gravity, the evolution of 
$a(t)$ is determined by the following system of equations,
\begin{equation}
	H^2\equiv \Big(\frac{\dot{a}}{a}\Big)^2=\frac{8\pi G}{3}\Big[\rho_\phi +\rho_m \Big ]\, ,
	\label{F1}
\end{equation}
\begin{equation}
	\frac{\ddot{a}}{a}=-\frac{4\pi G}{3}\Big [\rho_m +\rho_\phi +3(p_m+p_\phi )\Big ]\, ,
	\label{F2}
\end{equation}
where an over-dot represents derivative with respect to the comoving time $t$ and $\rho_{\phi}$
and $\rho_m$ represent the scalar field and matter density respectively. 
The time evolution of $\phi(t)$ couples to $a(t)$ and $V(\phi)$ and is governed by the second 
order generally non-linear  differential equation, 
\begin{equation}
	\ddot{\phi}+3H\dot{\phi}+\frac{dV}{d\phi}=0\, . 
	\label{continiuty}
\end{equation}  
We also have the continuity equation which holds separately for each component,
\begin{equation}
	\dot{\rho}+3H \rho\bigl(1+\omega\bigr)=0 \, ,
	\label{cequation}
\end{equation}
where ${\omega} \equiv p/\rho$ represents the equation of state parameter for the component 
under consideration. The fluids filling the universe have equation of state given by,
\begin{equation}
	p_m=0   , \quad p_\phi=\omega_\phi \rho_\phi \, ,
\end{equation}
with 
\begin{equation}
	\rho_m=\frac{\rho^0_{m}}{a^3}~\footnote{In this article, we use super or subscript 0 to represent the 
		values of the dynamical variables at present time $t_0$ with the convention $a_0 = 1$, 
		so that present day redshift $z_0 = 0$}, \quad  \rho_{\phi} =\frac{\dot{\phi^2}}{2}+V, \quad p_\phi =\frac{\dot{\phi^2}}{2}-V\, .
	\label{phipro}
\end{equation}
Using $p_{\phi}$ and $\rho_{\phi}$ from Eqn.~\ref{phipro} in Eqn.~\ref{cequation} 
along with $p_{\phi} =\omega_{\phi} \rho_{\phi}$, we can rewrite continuity equation as follows,
\begin{equation}
	\frac{d}{da}\Big (\frac{\dot{\phi^2}}{2}\Big)+\frac{dV}{da}+\frac{3}{a}\Big (\frac{\dot{\phi^2}}{2}\Big )=0\, .
\end{equation}
Now, by considering $V=V(a)$ and using the transformation $\dot{\phi^2}(a)=y(a)$  we  get the following 
linear differential equation,
\begin{equation}
	\frac{dy}{da}+2\frac{dV}{da}+\frac{6}{a}y=0\, ,
	\label{MasterEqn}
\end{equation}
where we have assumed that  $y$ and $V$ are in units of present day critical energy density $\rho^0_{c}$ such that,
\begin{equation}
	y\rightarrow y'=\frac{y}{\rho^0_{c}}, \quad V\rightarrow V'=\frac{V}{\rho^0_{c}}\, ,
	\label{normalization}
\end{equation}
and omitted any primes in our forthcoming analysis of this work. The solution of  
linear differential equation~\ref{MasterEqn}  gives  $\dot \phi^2(a)$~\footnote{We note that 
	with the aforementioned normalization choice both $V$ and $\dot \phi^2$ henceforth become 
	dimensionless variables.}. Using Eqn.~\ref{normalization} the Friedmann Eqns.~\ref{F1} and~\ref{F2} 
become, 
\begin{equation}
	H^2(a) = H^2_0\left(\frac{\dot \phi^2}{2} + V(a) + \frac{\Omega^0_m}{a^3}\right) \, ,
	\label{F3}
\end{equation}
and 
\begin{equation}
	\frac{\ddot a}{a} =   -H^2_0\left({\dot \phi^2} - V(a) + \frac{\Omega^0_m}{2a^3}\right) \, ,
	\label{F4}
\end{equation} 
where we have assumed $p_m = 0$. 
At this point, we mention that once a form of $V(a)$ is assumed, Eqn.~\ref{MasterEqn} may be considered 
to be  decoupled from the expansion dynamics of the universe, i.e., it does not depend on the exact 
solution of $a(t)$ which is obtained from first and second Friedman equations, Eqns.~\ref{F3} and~\ref{F4}.
This is an excellent advantage since the independent variable $y(a)$ can now be evolved irrespective of 
evolution of other components that contribute to the total energy momentum tensor. Another advantage of 
Eqn.~\ref{MasterEqn} is that  using its solution we immediately obtain the equation of 
state parameter of scalar field as, $\omega_\phi $ as 
\begin{equation}
	\omega_{\phi}(a)=\frac{{\dot{\phi^2}}/{2}-V}{{\dot{\phi^2}}/{2}+V} \, ,
	\label{omega_phia}
\end{equation}
which serve as an important physical parameter to describe and constrain the nature of dark energy
from both observational and theoretical point view. In this work, we focus ourselves for cases where 
both $\dot \phi^2 $ and $V(a)$ are positive (semi) definite,  resulting 
in $-1 \leq \omega_\phi(a) \leq 1$. 

Once $\dot \phi^2(a)$ is solved using Eqn.~\ref{MasterEqn} for the assumed model of $V(a)$  one 
can relate the dynamics of the $\phi$ sector with some important dynamical variables of the dynamics 
of $a(t)$. One such variable is  the  total effective equation of state
parameter, $\omega_{\textrm tot}$ taking into account all components of the universe, 
\begin{equation}
	\omega_{\textrm{tot}}(a)\equiv \frac{p_\phi + p_m}{\rho_\phi + \rho_m} \equiv 
	\frac{{\dot{\phi^2}(a)}/{2}-V(a)}{{\dot{\phi^2}(a)}
		{/2}+V(a)+{\Omega^0_{m}}/{a^3}}\, ,
	\label{omega_phit}
\end{equation}
where $\Omega^0_{m}$ represents the present day energy density parameter for matter. 
The effective equation of state parameter, $\omega_{\textrm{tot}}$  has great significance because it tells us whether 
the universe undergoes an accelerated expansion ($\omega_{\textrm{tot}}<-1/3$) or decelerating expansion ($\omega_{\textrm{tot}}>-1/3$)
at any particular epoch of time. Another useful parameter which encodes the information of acceleration or deceleration 
phases of $a(t)$ in its sign is the so-called deceleration parameter, $q$, defined as,
{\begin{equation}
		q=-\frac{\ddot{a}}{\dot{a^2}}a= -\frac{\ddot{a}}{a}\Big (\frac{\dot{a}}{a}\Big )^{-2}
		=    -\frac{\ddot{a}}{a}\big (H \big)^{-2} \, .
		\label{dec_par}
	\end{equation}
	A transition between the two phases always corresponds to zeros of the $q$ parameter. 	
	Since $\dot{\phi^2}(a)$, $V(a)$ and $\rho_m(a)$ are now known functions of scale factor, 
	by using Eqn.~\ref{F1} we can easily obtain  the Hubble parameter, $H(a)$~\footnote{Solutions corresponding 
		to  $H>0$ correspond to expanding cosmological models and $H<0$ corresponds to collapsing models.}. Knowing
	the numerator of Eqn.~\ref{dec_par} using Eqn.~\ref{F4} and denominator using Eqn.~\ref{F3}  we find 
	an exact expression of $q(a)$ as well.

	Before we proceed to discuss solution of Eqn.~\ref{MasterEqn} for specific choice of $V(a)$ 
	let us discuss some general features of these solutions. First, $a =0$ is a singular point 
	of Eqn.~\ref{MasterEqn}. This means that the solution $\dot \phi^2(a)$ is not analytic at $a = 0$. 
	One common feature of $\dot \phi ^2$ may be obtained by considering the associated homogeneous 
	equation corresponding to Eqn.~\ref{MasterEqn} by setting either $V = 0$ or $V = V_0$, a constant for
	all $a$, so that $dV/da = 0$. In this case, ignoring the trivial solution, $\dot \phi^2(a) = 0$, 
	we have $\dot \phi ^2(a) = \dot \phi^2_0/a^6$. In fact, presence of a  modulation factor of 
	$\sim 1/a^6$ is ubiquitous in the solution of $\dot \phi^2(a)$ through the integrating factor 
	of Eqn~\ref{MasterEqn} for any other choice of $V(a)$. Thus $\dot \phi^2$ diverges in general, 
	as $a \rightarrow 0$.  This can be contrasted with  the corresponding behavior
	for matter (or radiation) density, $\rho_m(a) = \rho^0_m/a^3$ ($\rho_r(a) = \rho^0_r/a^4$).
	We note that, such divergence of $\dot \phi^2(a)$ does not necessarily imply divergence of 
	$\omega_\phi$, for a general $V(a)$. If $V = V_0$ we have $\omega_\phi(a) =\left(\dot \phi^2_0 
	- 2a^6V_0\right)/\left(\dot \phi^2_0 + 2a^6V_0\right)$, which tends to unity as $a \rightarrow 0$
	~\footnote{Apart from the usual argument that the classical Friedmann equation must only be 
		valid up to some  initial scale factor well above the Planck length scale, the problem of 
		divergence of $\dot \phi^2$ can also be bypassed if we assume that the scalar field theory valid 
		up to some initial scale factor $a_i$ corresponding to $\dot \phi^2(a_i) = \dot \phi^2_i$, 
		a finite value.}.
	
	
	\section{Applications}
	\subsection{Power Law Potential}
	\label{sec:power_law}
	
	Let us consider the power law form of potential energy density,
	\begin{equation}
		V(a)=V_{0} a^n \, ,
		\label{pl_Va}
	\end{equation} 
	where $V_{0}$ is a dimensionless constant and $n$ is a real number. By direct observation of 
	Eqn.~\ref{MasterEqn} we see that in this case, $\dot \phi^2(a) \sim a^n$ is  a solution. If 
	$n > 1$ all terms of Eqn.~\ref{MasterEqn} remains finite for all $a$.  In fact, for any polynomial
	choice of 
	\begin{equation}
		V(a) = \sum_{p=0}^nV_pa^{p+1} \, ,
	\end{equation}
	where $V_p$ for $p =1, ...., n$ are constants, $\dot \phi^2(a)$ also admits a polynomial solution
	\begin{equation}
		\dot \phi^2(a) =  \sum_{p=0}^nA_pa^{p+1} \, , 
	\end{equation}
	which is finite for all $a$. In this case,~\ref{MasterEqn} becomes, 
	\begin{equation}
		\sum_{p=0}^nA_p\left(p+7 \right)a^p = -2\sum_{p=0}^nV_p(p+1)a^p\, , 
	\end{equation}
	which leads to  a solution of the form 
	\begin{equation} 
		\dot \phi^2(a) = -2\sum_{p=0}^n \frac{V_p\left(p+1\right)}{\left(p+7\right)}a^{p+1} \,. 
		\label{series_sol}
	\end{equation} 
	Eqn.~\ref{series_sol}  does not capture the complete picture of the most general 
	solution $\dot \phi^2(a)$ since it only corresponds to the solution of the in-homogeneous 
	Eqn.~\ref{MasterEqn}. The general solution of Eqn.~\ref{MasterEqn} is obtained after adding
	with Eqn.~\ref{series_sol} $C/a^6$, which is the solution corresponding to the 
	homogeneous equation of Eqn.~\ref{MasterEqn}, where $C$ is a constant. The 
	general solution is therefore, 
	\begin{equation}
		\dot \phi^2(a) = -2\sum_{p=0}^n \frac{V_p\left(p+1\right)}{\left(p+7\right)}a^{p+1} + \frac{C}{a^6}\,.
		\label{series_sol1}
	\end{equation}
	

	The constant $C$ is fixed by imposing suitable  boundary condition. If the potential energy density 
	consists of a single power law as given by Eqn.~\ref{pl_Va}, the only value of $p$ in Eqn.~\ref{series_sol1}
	becomes $p = n -1$.  Eqn.~\ref{series_sol1} now becomes, 
	\begin{equation}
		\dot{\phi^2}(a)\equiv y(a)= \frac{1}{a^6}\Big (-\frac{2V_{0} n a^{n+6}}{n+6}+C \Big)\, ,
		\label{phi2_1}
	\end{equation}
	where 
	$n \ne -6$. If at $t = t_0$, $a=a_{0}=1$ and $\dot{\phi^2}(1)=y_{0}$, then
	$\rho^0_{\phi} +\rho^0_{m} = \rho^0_{c} = 3H^2_0/(8\pi G)$, which following the definitions of Eqn.~\ref{normalization} 
	implies,  
	\begin{equation}
		{y_0} = 2\bigl(1-V_0-\Omega^0_{m}\bigr)\, .
		\label{c1}
	\end{equation}
	Using Eqns.~\ref{phi2_1} and ~\ref{c1} we obtain,
	\begin{equation}
		C= y_{0}+\frac{2V_{0}n}{n+6} = 2\biggl[\frac{n+6\left(1-V_0\right)}{n+6}\biggr] - 2\Omega^0_m\, , 
		\label{const_value}
	\end{equation}
	Finally using Eqns.~\ref{phi2_1} and~\ref{const_value} we obtain, 
	\begin{equation}
		\dot{\phi^2}(a) =  \frac{1}{a^6}\Bigg [2 \biggl(\frac{n+6\left(1-V_0\right)}{n+6}\biggr)
		-\frac{2V_{0} n a^{n+6}}{n+6} - 2\Omega^0_m\Bigg]\, .
		\label{phi2_2}
	\end{equation}

	To get some insight into the nature of the solution given by Eqn.~\ref{phi2_2}  we first 
	consider a simple case for which $n = 0$ (implying $V = V_0 $, a constant) and $\Omega^0_m = 0$ 
	as well. This corresponds to a dark energy dominated flat universe with $\dot{\phi^2}_0 = 2(1-V_0)$. 
	If $V_0 \ne 1$, $V \ne 1$ for always and using Eqn.~\ref{phi2_2},
	\begin{eqnarray}
		\dot \phi^2(a) = \frac{2(1-V_0)}{a^6}\, ,
		\label{phi2_3}
	\end{eqnarray}
	implying the kinetic energy density of the scalar field decays as $\sim a^{-6}$ as the universe 
	expands. One can further consider two  limiting values of $V_0$, namely, $1$ and $0$ respectively. 
	If  $V_0 =1$ then from Eqn.~\ref{phi2_2} $\dot \phi^2(a) = 0$ for all $a$ implying $\phi$ is a 
	constant. In this case, as expected, we recover the cosmological model with cosmological constant 
	with $\omega_{\phi}(a) = -1$. If $V_0 =0$, $\dot \phi^2_0 = 2$ to satisfy the flatness condition 
	today.  In this case, $\omega_{\phi} = 1$ always and hence the particular scalar field theory does 
	not cause inflation. Using Eqn.~\ref{phi2_3} we obtain $\dot \phi^2(a) = 2/a^6$. Substituting this 
	in first Friedmann equation with $\Omega^0_m = 0$ and after some algebra one finds, 
	\begin{equation}
		a(t) = \big(H_0\big)^{1/3}t^{1/3}\, ,
		\label{a_t}
	\end{equation}
	where we have used boundary condition $a = 0$ at $t=0$. Clearly this particular 
	scalar field does not cause inflation as argued earlier since $\omega_{\phi} =1$. 
	In summary, we conclude that for different choices of numerical 
	values of $V_0$ and specific choice of exponent $n$ in Eqn.~\ref{pl_Va} the 
	solutions $\dot \phi^2(a)$ as given by Eqn.~\ref{MasterEqn} are capable of 
	capturing a wide range of dynamical behavior of scale factor $a(t)$.   
	
	Using the general solution $\dot \phi^2(a)$ (Eqn.~\ref{phi2_2}) and Eqn.~\ref{pl_Va} 
	in Eqn.~\ref{omega_phia} and Eqn.~\ref{omega_phit} respectively, we obtain,
	\small
	\begin{eqnarray}
		\omega_{\phi} = \frac{n +6(1-V_0) - \Omega^0_m(n+6) - 2V_0(n+3)a^{n+6}}
		{n +6(1-V_0) - \Omega^0_m(n+6) +6V_0a^{n+6}}\, ,
	\end{eqnarray}
	\normalsize
	and
	\small
	\begin{eqnarray}
		\begin{split}
			\omega_{\textrm{tot}} = \Bigg(n +6(1-V_0) - \Omega^0_m(n+6) - 2V_0(n+3)a^{n+6}\Bigg)\Bigg/\\
			\Bigg(n +6(1-V_0) - \Omega^0_m(n+6) +6V_0a^{n+6} + \Omega^0_ma^3(n+6)\Bigg)\, . 
		\end{split}
	\end{eqnarray}

	\normalsize
	One can also easily calculate the Hubble parameter $H(a)$ as a function of scale 
	factor using the Eqn.~\ref{F3},
	\small
	\begin{equation}
		\begin{split}
			H^2(a)=H_0^2 \Biggl[ \frac{1}{a^6}\Biggl(\frac{n+6(1-V_0)}{n+6}-
			\Omega^0_m-\frac{V_{0} n a^{n+6}}{n+6} \Biggr)\\+ 
			V_{0} a^n+\frac{\Omega_{m0}}{a^3} \Biggr ]
		\end{split}	
		~\footnote{$H>0$ correspond to expanding cosmological models and $H<0$ 
			corresponds to collapsing models.}\,
	\end{equation}
	
	\normalsize
	
	The deceleration parameter $q(a)$ can be obtained from the Eqn.~\ref{dec_par}
	\small
	\begin{equation}
		\begin{split}
			q(a) = \Bigg(\Omega^0_m(a^3-4)(6+n)+n(4-6a^{n+6}V_0)\\-12(-2+(2+a^{n+6})V_0)\Bigg)\Bigg/\\\Bigg(2(6+n+\Omega^0_m(n+6)(a^3-1)-6 V_0+6a^{n+6}V_0)\Bigg)\, .
		\end{split}
	\end{equation}
	\normalsize

	\section{Reconstruction of Scalar Field Potential Energy}\label{sec:reconstruct}
	The first order differential Eqn.~\ref{MasterEqn} has another great advantage. It can be 
	used as a tool to reconstruct $V(a)$ and $\dot \phi^2(a)$ from any general equation of state 
	parameter, $w_\phi(a)$ of the scalar field. To illustrate this, we first note that,  
	\begin{eqnarray}
		{\dot \phi^2}(a) = \biggl[\frac{\omega_\phi+1}{1-\omega_{\phi}}\biggr]\left(2V(a)\right)\, , 
		\label{phid2}
	\end{eqnarray} 
	where we have used Eqn.~\ref{omega_phia}~\footnote{In this case, we have assumed 
		that $\omega_\phi(a) \ne 1$  or equivalently, $V(a) \ne 0$  
		for the domain of interest of $a$. The solution $V(a) = 0$  when
		$\omega_\phi(a) = 1$ can be easily obtained from Eqn.~\ref{omega_phia}.} After some algebra, we obtain,
	\begin{eqnarray}
		\frac{d\dot\phi^2(a)}{da} = \frac{2V}{\left(1-\omega_\phi\right)^2}\frac{d\omega_\phi}{da}
		+ 2\frac{dV}{da}\biggl[\frac{1+\omega_\phi}{1-\omega_\phi}\biggr]\, .
		\label{dphid2da}
	\end{eqnarray}
	Using Eqns.~\ref{phid2} and~\ref{dphid2da} in Eqn.~\ref{MasterEqn}, after some algebra we 
	obtain,
	\begin{eqnarray}
		\frac{dV}{da} + V\biggl[\frac{3 \left(1 + \omega_\phi\right)}{a} + 
		\frac{1}{2\left(1-\omega_\phi\right)}\frac{d\omega_\phi}{da}\biggr] = 0\, , 
		\label{V}
	\end{eqnarray}
	which  has the solution, 
	\small
	\begin{eqnarray}
		V(a) = V_0 \exp\biggl[ -\int^a_{1}
		\biggl\{\frac{3 \left(1 + \omega_\phi\right)}{a} + \frac{1}{2\left(1-\omega_\phi\right)}
		\frac{d\omega_\phi}{da}\biggr\}da\biggr] \, ,  
		\label{V1}
	\end{eqnarray}
	\normalsize
	where $V_0 = V(a = a_0 = 1)$ as earlier. It is evident from Eqn.~\ref{V1} that given any general 
	form of $\omega_\phi$ an analytical 
	expression of scalar field potential energy density can be obtained as long as the first and 
	second integrands in the exponent of this equation are integrable. The second integral
	in the exponent can further be carried out analytically.  Performing integration by parts 
	on the first integrand,  and after some algebra as in Appendix we find, 
	\begin{equation}
		\begin{split}
			V(a) = V_0\biggl[\frac{1-\omega_\phi(1)}{1-\omega_\phi(a)}\biggr]^{1/2}a^{-3\left(1+\omega_\phi(a)\right)}\\
			\times\exp\biggl\{3\int_1^a\frac{d\omega_\phi}{da}\ln(a)da\biggr\}\, ,			
		\end{split}
		\label{V2}
	\end{equation}
	\normalsize
	It is interesting to note from Eqn.~\ref{V2} that apart from  the scale factor dependence through 
	$\omega_\phi$ induced by the first term  $\bigl[(1-\omega_\phi(1))/(1-\omega_\phi(a))\bigr]^{1/2}$,
	$V(a)$ is determined by the product of one power law and another exponential function in $a$.  As a simple application of
	Eqn.~\ref{V2} if we assume $\omega_\phi = -1$ for all $a$, the exponential function on $a$ becomes 
	unity  and since $d\omega_\phi/da =0$, the integral in the exponent disappear as well, implying $V = V_0$, a constant.
	In this case, $\dot \phi^2 = 0$ by using Eqn.~\ref{phid2}, as expected. Therefore, starting from
	$\omega_{\phi} = -1$ we  
	reproduce the well known case of a cosmological constant by using the general solution of $V(a)$
	(Eqn~\ref{V2}) and definition of $\omega_\phi$ (Eqn.~\ref{phid2}). If we assume $\omega_\phi(a) =0$ for all $a$,
	then using Eqn.~\ref{V2} we obtain $V(a) = V_0/a^3 $ and $\dot \phi^2/2 = V(a)$ using Eqn~\ref{omega_phia}, 
	i.e. the kinetic energy of the scalar field traces the potential energy exactly for all $a$.
	In the case, the scalar field behaves like pressure less non relativistic matter. 
	One can also reconstruct the scalar field potential from the observed spectral indices for the density perturbation $n_s$ and the tensor to scalar ratio $r$ as seen in~\citep{2016arXiv161106680B}. 
	
	An interesting application of Eqn.~\ref{V1} is to obtain potential energy density for the 
	CPL equation state 
	\begin{eqnarray}
		\omega_\phi(a)  = \omega_0 + \omega_1a \, ,
		\label{CPL1}
	\end{eqnarray}
	where $ \omega_0$ and  $\omega_1$ are constants~\footnote{In fact, the CPL equation 
		of state parameterized by, $ \omega_\phi(z)  = \omega_a + \omega_b\frac{z}{1+z}$, where $z$ denotes 
		the redshift. Using $1/a = 1+z$ and a some algebra  with 
		some redefinition of variables, one can show this is equivalent to Eqn.~\ref{CPL1}}. Using Eqn.~\ref{CPL1}
	in Eqn.~\ref{V2}  we obtain, 
	\begin{equation}
		V(a) = V_0\biggl[\frac{1-\omega_0-\omega_1}{1-\omega_0-\omega_1a}\biggr]^{1/2}a^{-3\left(1+\omega_0 + \omega_1a \right)}
		a^{\omega_1}\, .	
		\label{CPL_V}
	\end{equation}
	If $\omega_1 = 0$ in Eqn.~\ref{CPL1} so that $\omega_\phi = \omega_0 \ne 1$, Eqn~\ref{CPL_V} leads to
	\begin{eqnarray}
		V(a) = V_0a^{-3\left(1+\omega_0\right)}\, .
		\label{CPL_V1}
	\end{eqnarray}
	By using Eqn.~\ref{omega_phia} we can also obtain,
	\begin{equation}		
		\dot \phi^2(a) = 2V_0^{-3\left(1+\omega_0\right)}\left(1+\omega_0\right)/\left(1-\omega_0\right). 
	\end{equation} 
	
	\section{Relation with the Particle Physics Models}\label{sec:physical}
	Once we obtain $\dot \phi^2(a)$ analytically, it is possible to get an analytical expression 
	for the Hubble parameter $H(a)$ using Eqn.~\ref{F3}. Then by a numerical evaluation one can 
	easily obtain $\phi(a)$ and finally the particle physics motivated form i.e. potential in 
	terms of  the scalar field, $V(\phi)$. This can be achieved by the following procedure. It is 
	simple to estimate $\dot a = aH(a)$, where we already obtained an analytical expression for 
	the Hubble parameter $H$  following our method. Using identity 
	$\dot a = (da/d\phi)\dot \phi$ and knowing $\dot \phi$ following the method of this article, 
	one obtains $\phi(a)$ after employing a numerical
	integration. From the graph of $\phi (a)$ and $V(a)$ one can readily estimate $V(\phi)$ numerically. This
	makes a connection of our method with the theory of particle physics, since there $V(\phi)$ is 
	directly model led following symmetry conditions of the theory. Thus for any choice of potential $V(a)$, one can obtain its corresponding form of $V(\phi)$. 
	For instance, considering the case of the potential $V(a)$ in the power law form discussed in 
	Sec.~\ref{sec:power_law}, one can easily obtain the corresponding  form of $V(\phi)$ using the 
	method mentioned above. Here the scalar field $\phi$ is scaled by $\sqrt{8\pi G}= M_p^{-1}$. 
	The Fig.~\ref{Vphi} shows the variation of scalar field potential as a function of scalar field 
	for the case of the potential $V(a)=V_0 a^n$  with $V_0=0.672$ and $n=-0.24$. It is very interesting
	to note that, this potential approximates  an inverse power law potential of the form
	$V(\phi)=V^\prime_0 (\phi+ \phi_0)^{-m}$ with $V^\prime_0= 0.16$, 
	$\phi_0=0.00049$ and  $m= 0.257$ with some level of clearly visible differences. This is shown in blue 
	in Fig.~\ref{Vphi}.  It is noteworthy to mention that the derived potential $V(\phi)$ 
	resembles to the class of quintessential tracker poentials (inverse power law models) proposed by Ratra 
	and Peebles in 1998.~\citep{1988PhRvD..37.3406R}.
	
	\begin{figure}[ptb]
		\includegraphics[height=6cm]{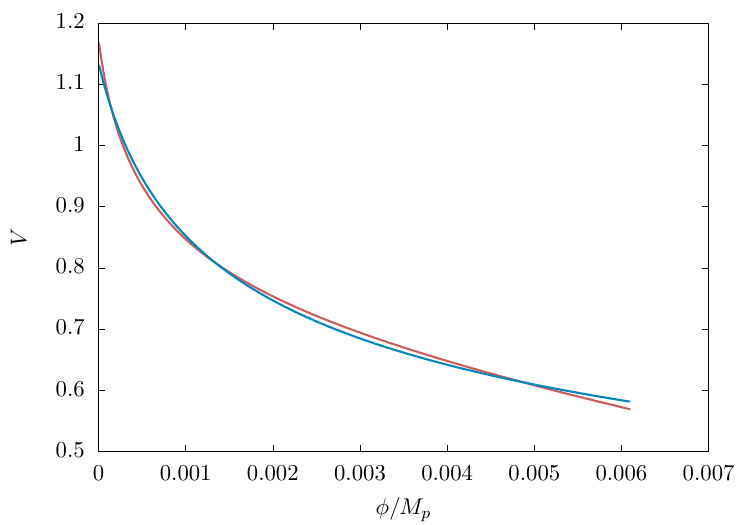}
		
		\caption{Figure showing the reconstructed potential $V(\phi)$ (in red)  with $\phi$  
			when  $V(a)=V_0 a^n$. The blue curve shows the  potential $V(\phi)=V^\prime_0 (\phi+ \phi_0)^{-m}$ with $V^\prime_0= 0.16$, 
			$\phi_0=0.00049$ and  $m= 0.257$.
		}
		\label{Vphi}
	\end{figure}

	\section{Observational Constraints}
	\label{sec:constraints}
	
	After obtaining the analytical expressions for Hubble parameter as a function of redshift (or scale factor) for different cosmological models, one can constrain its model parameters using the observational probes in the cosmology. The observational probes utilized for the data analysis includes the Type 1a Supernovae, Baryon Acoustic oscillations and Hubble parameter data sets. We constrain the parameters by employing the $\chi^2$ statistics. The total likelihood function for the joint data is given by,
	\begin{equation}
		\mathcal{L}_{tot}(\theta) = e^{- \chi_{tot}^2(\theta)\big /2}
		\label{liklie}
	\end{equation}
	where
	\begin{align}
		\chi_{\rm tot}^2=\chi_{\rm SN}^2+\chi_{\rm Hub}^2+\chi_{\rm BAO}^2 \,,
		\label{chi_tot}
	\end{align}
	
	where $\theta$ denotes the parameters of the model under consideration. The best fit values of the model parameters($\theta$) are the values corresponding to the minimum $\chi^2$ value.
	\subsection{\textbf{Observational Probes}}
	\subsubsection{\textit{Hubble Data (H).}}
	The Hubble parameter data set consists of measurements of the Hubble parameter $H(z)$ at different redshifts. We use $H(z)$ data which is compiled and listed in Table-1 of \citep{2017ApJ...835...26F}. The table contains 38 measurements of Hubble parameter and its associated errors in measurements up to a redshift of z= 2.36. From the total of 38 data sets, we have considered only 32 points as we do not consider three data points taken from Alam et. al.(2016) at reshifts z= 0.38, 0.51, 0.61 and we also removed the data points corresponding to the redshits z= 0.44, 0.6, 0.72 as they are already included in the BAO dataset. 
	The chi-squared function of Hubble data is given by,
	\begin{align}
		\chi_{\rm Hub}^2(\theta)=\sum_{i=1}^{32}
		\frac{\left[H_{\rm th}(z_i,\theta)-H_{\rm obs}(z_i)\right]^2}{
			\sigma^2_H(z_i)}\,,
	\end{align}
	were  $\sigma_H$ is the error associated with each measurements.

	\subsubsection{Type Ia Supernovae (SN Ia).}
	
	The Type 1a supernovae are the result of the explosion of a white dwarf star in a binary
	when it crosses the Chandrasekhar limit. These Type 1a supernovae is an ideal probe for the
	study of the cosmological expansion. As they all have the same luminosity, they are considered
	as a good standard candle. So the first data set which we used for the analysis is the
	Type 1a Supernovae data from the Pantheon compilation \citep{2018ApJ...859..101S}. This data set consists of 1048
	supernovae in the redshift range $0.01 < z <2.26$ .

	The luminosity distance of a Type 1a supernova at given redshift $z$ reads as,
	\begin{equation}
		D_L(z)=(1+z)
		\int_0^z\frac{H_0dz'}{H(z')}.
	\end{equation}
	Moreover the luminosity distance is directly related to the observed quantity, distance modulus $\mu(z)$ given by,
	\begin{equation}
		\mu(z)=m - M=5 \log D_L(z)+\mu_0
	\end{equation}
	
	where $M$ and $m$ are the absolute and apparent magnitude of the supernovae. Here the quantity $\mu_0=5 \log\left(\frac{H_0^{-1}}{\rm Mpc}\right)+2 5$ is a nuisance parameter which should be marginalized.
	
	So for the case of SN1a, the $\chi^2_{SN}$  estimator is defined as, 
	\begin{align}
		\chi_{\rm SN}^2(\mu_0,\theta)=\sum_{i=1}^{1048}
		\frac{\left[\mu_{th}(z_i,\mu_0,\theta)-\mu_{obs}(z_i)\right]^2}{
			\sigma^2_\mu(z_i)}\,,
	\end{align}
	
	where $\mu_{th}$, $\mu_{obs}$ and $\sigma_\mu$ are the theoretical, observed distance modulus and the unertainty in the observed quantity respectively. Here $\theta$ represents the parameters of the model under consideration. After marginalizing $\mu_0$ and by following the reference \citep{2005PhRvD..72l3519N}, we get

	\begin{equation}
		\chi_{\rm SN}^2(\theta)=A(\theta)-\frac{B^2(\theta)}{C(\theta)}\,,
	\end{equation}
	
	where,
	
	\begin{equation}
		\,\,A(\theta) =\sum_{i=1}^{1048}
		\frac{\left[\mu_{th}(z_i,\mu_0=0,\theta)-\mu_{obs}(z_i)\right]^2}{
			\sigma^2_\mu(z_i)}\,, \\
	\end{equation}
	
	\begin{align}
		&B(\theta) =\sum_{i=1}^{1048}
		\frac{\mu_{th}(z_i,\mu_0=0,\theta)-\mu_{obs}(z_i)}{\sigma^2_\mu(z_i)}\,, \\
		&C(\theta) =\sum_{i=1}^{1048} \frac{1}{\sigma^2_\mu(z_i)}\,.
	\end{align}

	\subsubsection{Baryon Acoustic Oscillations (BAO).}
	
	The Baryon Acoustic Oscillations (BAO), which are considered as the standard rulers of the cosmology, are frozen relics left over from the pre-decoupling universe. Here we have used the BAO data from 6dFGS, SDSS DR7 and WiggleZ at redshifts z = 0.106, 0.2, 0.35, 0.44, 0.6 and 0.73. In order to derive the BAO constraints we make use of the Distance parameter $D_v(z)$ which is a function of angular diameter distance and Hubble parameter given by,
	\begin{equation}
		D_v(z) = \Big[(1+z)^2 d^2_A(z)\frac{z}{H(z)}\Big]^\frac{1}{3}
	\end{equation}
	Here $d_A(z)$ is the angular diameter distance. We use the measurements of acoustic prameter $A(z)$ from \citep{2011MNRAS.418.1707B}, where the theoretically predicted $A_{th}(z)$ is given by Eq. 5 of \citep{2005ApJ...633..560E}.
	\begin{equation}
		A_{th}(z) = \frac{100 D_v(z) \sqrt{\Omega_{m0}h^2}}{z}
	\end{equation}
	
	After following the procedures given in Sec. 5.4 of \citep{2013arXiv1309.3710O}, one can obtain the acoustic parameter $A_{th}$ independent of the Hubble constant $H_0$. Finally after some algebra the chi-squared function of BAO data \citep{2011MNRAS.418.1707B} reads as,
	\begin{align}
		\chi_{\rm BAO}^2(\theta)=\sum_{i=1}^{6}
		\frac{\left[A_{th}(z_i,\theta)-A_{obs}(z_i)\right]^2}{
			\sigma^2_A(z_i)}\,,
	\end{align}
	
	
	\begin{center}
		
		\begin{table}
			\addtolength{\tabcolsep}{0.5cm}
			\renewcommand{\arraystretch}{1.3}
			\centering
			\begin{tabular}{|c|c|}
				\hline
				\text{Parameter}           &\text{Prior}            \\
				\hline 				
				$\Omega_{m0}$       &[0, 1]          \\
				$V_0$       &[0, 1]          \\
				$n$               &[-5, 5]        \\
				$H_0$          &[55, 80]     \\
				
				
				\hline
			\end{tabular}
			\caption{Priors used for the MCMC analysis of the power law potential, $V_0 a^n$.}
			\label{tab:prior_power}
		\end{table}
	\end{center}		
	
	\subsection {\bf Methodology}\label{method}
	
	We used the Markov Chain Monte Carlo (MCMC) method to find the high confidence regions of the model parameters given a set of observational data .  We perform a likelihood analysis to minimize the $\chi^2$ function in Eq.  \ref{chi_tot} and thereby
	obtain the best-fit model parameters corresponding to the minimum $\chi^2$ value. The minimization of the $\chi^2$ is equivalent to the maximization of the likelihood function in Eq. \ref{liklie}. Here we constrain the  parameters of the power law potential given in Sec.~\ref{sec:power_law}. The model parameters and their prior values considered for the
	MCMC analysis is given in Table \ref{tab:prior_power}. 
	
	Initially, we begin the MCMC analysis with the usual way where we consider a simple proposal function of the form, $ \theta^j_{i+1}  = \theta^j_i  + \delta\theta^j \eta^j$, where $j$ is a parameter index, $\delta^j$ is a predefined rms step size, and $\eta^j$ is a Gaussian stochastic variate of zero mean and unit variance. As we are dealing with a model of five parameters, most of the parameters of interest will be strongly correlated and make this choice quite inefficient. So we performed certain optimization on the MCMC chain samples, which enables us to increase the acceptance ratio and also the chain convergence. After obtaining sufficient samples from the preliminary chain, we check the autocorrelation of the Markov chain. This will give us an idea for estimating how many iterations of the Markov chain are needed for effectively independent samples. Later we perform thinning on the initial chain we had already obtained so as to get less correlated samples. This action will further reduce the total number of samples.
	We will then compute the covariance matrix $C_{ij} = \big<\delta\theta^i\delta
	\theta^j\big>$of the resulting samples. Finally, we then Cholesky-decompose this matrix, $\bold C =\bold L \bold L^{t}$ where $\bold C$ is the covariance matrix, $\bold L$ and $\bold L^t$ are the lower triangular matrix and its conjugate transpose respectively. We then redefine our proposal function to be  $ \theta_{i+1}  = \theta_i  + \alpha \bold L \bold\eta$,  where $\eta$ is now a  vector of Gaussian variates and $\alpha$ is an overall scale factor, typically initialized at $ \sim 0.3$. This will help us to have the proposed samples with approximately correct covariance structure, and it will also improve the sampling efficiency significantly. In order to avoid very high and very small step sizes, we also impose a constraint that the acceptance ratio must be higher than $5\%$ and lower than $80\%$. One can adjust the scale factor $\alpha$ if one of these two criteria is violated.

	\begin{figure}[ptb]
		\includegraphics[height=8.5cm]{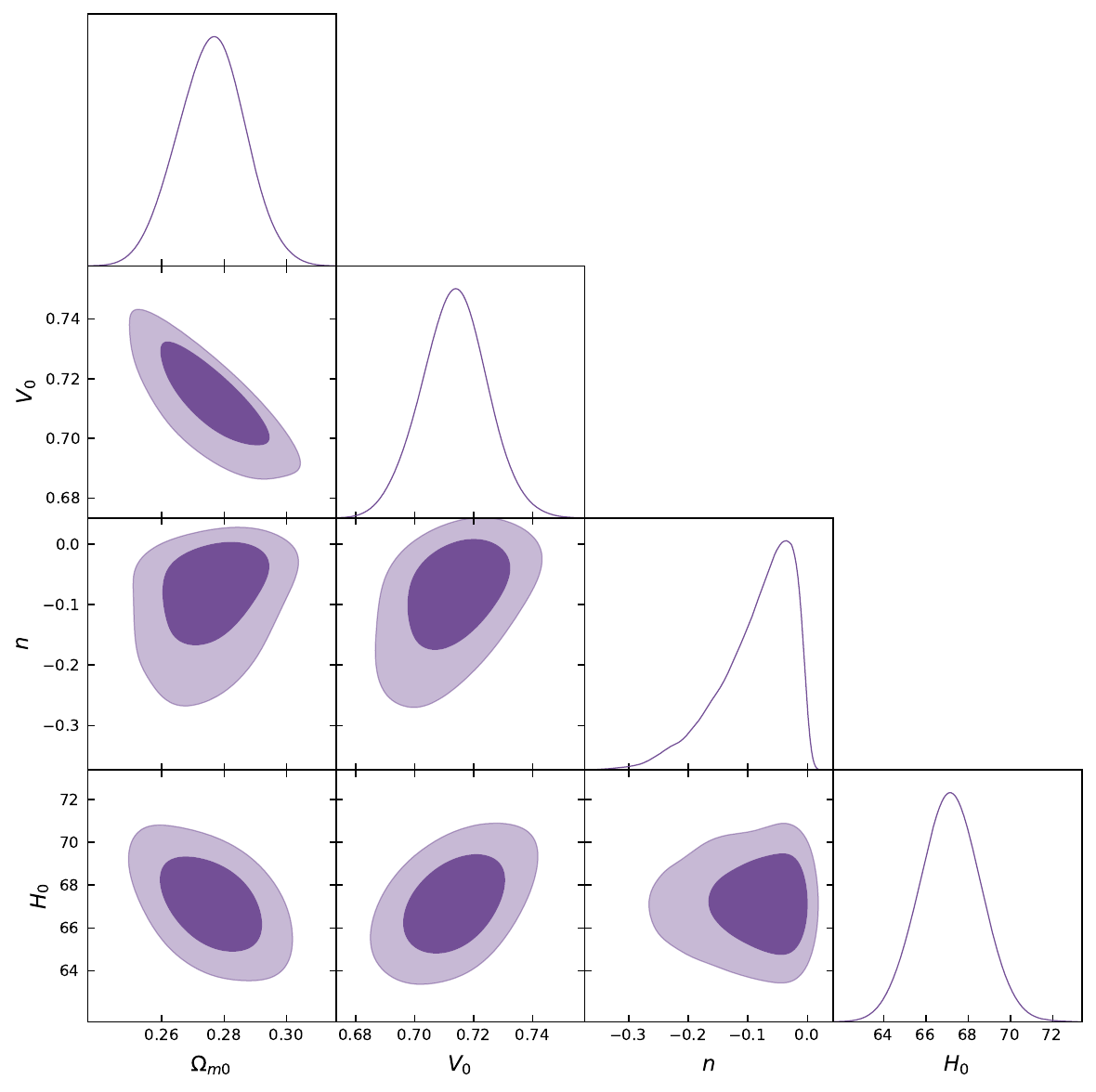}
		
		\caption{\small Posterior distribution for the model parameters ($\Omega_{m0}, V_0, n, H_0$) of the power law potential. The contours shows 68.3\% and 95.5\% confidence regions. The quantitative results are summerised in Table \ref{tab:best_fit_flat}.
		}
		\label{posterior}
	\end{figure}

	\begin{figure}[ptb]
			\vspace{-1.3em}
		\includegraphics[height=6cm]{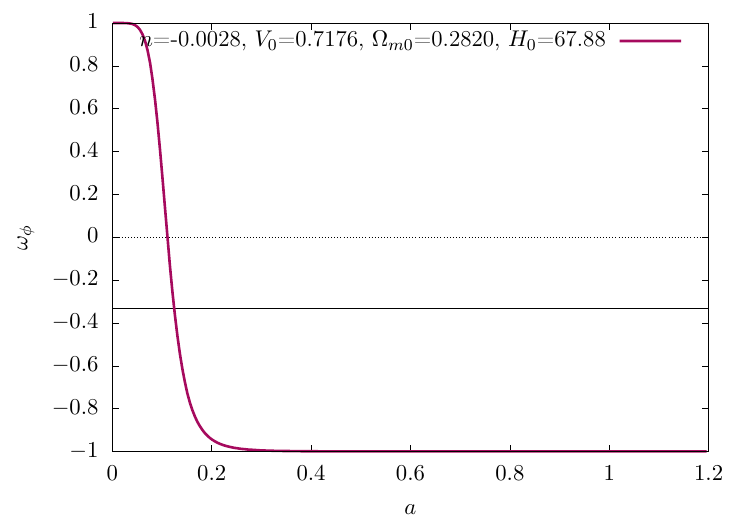}		
		\caption{\small Figure showing the variation of $\omega_{\phi}$ as a function of scale factor corresponding to the best-fit values of the power law potential. 
		}
		\label{wp_bestfit}
	\end{figure}
	
	\begin{figure}[ptb]
		\includegraphics[height=6cm]{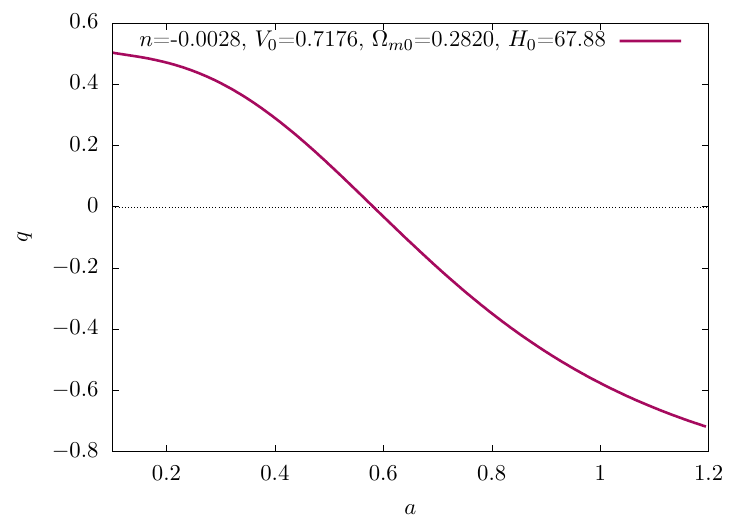}
		
		\caption{\small Figure showing the variation of deceleration parameter as a function of scale factor corresponding to the best-fit values of the power law potential. .
		}
		\label{de_bestfit}
	\end{figure}
	
	\begin{figure}[ptb]
		\includegraphics[height=6cm]{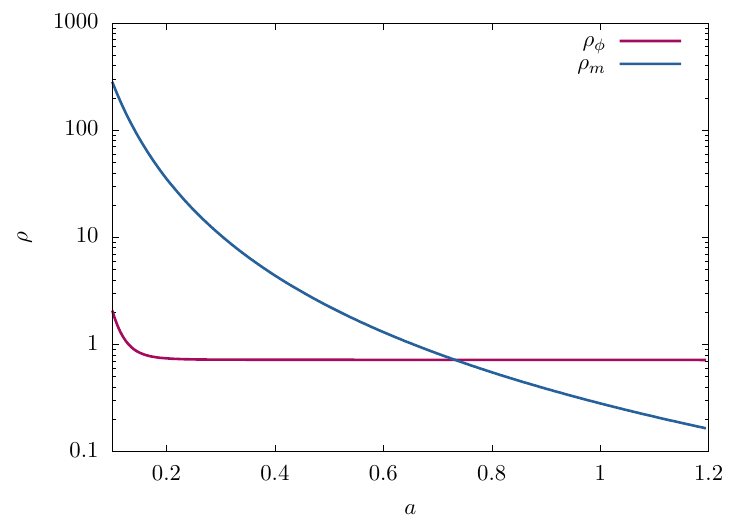}
		
		\caption{\small Figure showing the variation of energy density (log scale) of scalar field and non-relativistic matter as a function of scale factor corresponding to the best-fit values of the power law potential. .
		}
		\label{den_bestfit}
	\end{figure}
	
	\begin{table}[t]
		\addtolength{\tabcolsep}{1pt}
		\renewcommand{\arraystretch}{1.4}
		\centering
		\begin{tabular}{|l|c|c|}
			\hline 
			\text{Parameter} &  \text{Best-fit $\pm$ }   & \text{Mean $\pm$}\\
	     	\text{} &  \text{ 95.5\% limits}   & \text{ 95.5\% limits}\\
			\hline
			{$\Omega_{m0}$}       & $0.2820^{+0.029}_{-0.034}$      & $0.276 \pm 0.021$\\
			{$V_0        $}        & $0.7176^{+0.029}_{-0.038}$       & $0.713 \pm 0.022$\\
			{$n$}           & $-0.0028^{+0.006}_{-0.324}$       & $-0.085^{+0.090}_{-0.13}$\\
			{$H_0        $}        & $67.88^{+4.48}_{-4.60}$       & $67.1\pm 2.9$\\

			\hline
			\multicolumn{3}{|c|}{$\chi^2_{min} = 1054.0437$}\\
			\hline 
		\end{tabular}
		\caption{Best fit model  for the power law Potential.}
		\label{tab:best_fit_flat}
	\end{table}

	\section{Discussions and Conclusions }
	\label{sec:conclusion}

	The exact solutions of Einstein's equations play a very important role in cosmology as they
	help in the understanding of quantitative and qualitative features of the dynamics of the 
	universe as a whole. In this article, we discuss a method to obtain analytical expressions 
	for the equation of state parameter, deceleration parameter and Hubble parameter in spatially 
	flat FRW model of the universe with a perfect fluid and scalar field.
	
	\vspace{5pt}
	In Sec.~\ref{sec:basic} a method is proposed to  obtain analytical expressions for the kinetic 
	energy ($\dot{\phi^2}$) of the scalar field as a function of scale factor for different choices 
	of the scalar field potential. Once  the kinetic term ($\dot{\phi^2}$) is obtained, one can also 
	obtain exact solutions for  $\omega_{\phi}$, $\omega_{tot}$, $q$ and $H(a)$ by simple substitution. 
	The method proposed in this article can  also be applied to inflationary phase, to obtain analytical 
	solutions, where the scalar field dominates. The effective equation of state parameter $\omega_{tot}$ 
	and also the deceleration parameter are of great importance as they provide information about the 
	epochs of acceleration and deceleration phases of the universe.
	
	\vspace{5pt}
	It is important to emphasize the importance of exact expression of $H(a)$ (or $H(z)$) obtained 
	by us in scalar field cosmology with other perfect fluid components like matter and radiation. 
	The analytical results can be directly fitted with observations of $H(z)$ to estimate the best-fit
	values of parameters of the theory. Thus our method builds an important connection between theory and 
	observations to constrain the scalar field potential along with other cosmological parameters. 
	
	\vspace{5pt}
	In Sec.~\ref{sec:reconstruct}, we consider the inverse problem of 
	reconstructing the scalar field potential energy by assuming any general analytical expression of scalar field equation of state parameter as a function of the scale factor. By this method, we also reconstructed the scalar field potential for one of the most widely used parametrizations of dark energy called the Chevallier-Polarski-Linder (CPL) model.
	
	\vspace{5pt}
	Another important result that can be obtained using the results of this article, is to reconstruct the scalar field potential in terms of scalar field. In Sec.~\ref{sec:physical} we have discussed 
	the method  of reconstruction of $V(\phi)$ from the assumed 
	model of $V(a)$ and the exact analytical expressions of $\dot \phi(a)$, $H(a)$
	and using simple numerical integration. Thus we obtained that the potential $V(a)=V_0 a^n$ has a close resemblance to the class of quintessential tracker potentials proposed by Ratra and Peebles in 1988.
	
	\vspace{5pt}
	Finally in Sec. \ref{sec:constraints} we constrained the model parameters of the power law potential with the Hubble parameter data, Baryon Acoustic Oscillation(BAO) data and the recent Panthelon Type 1a supernovae Compilation. The contours showing the 68.3\% and 98.5\% confidence regions are depicted in Fig.~\ref{posterior}. The best-fit and the mean values are shown in Table~\ref{tab:best_fit_flat}. Interestingly, we observed that the $H_0 = 67.1\pm 1.5 \,Km/S/M_{pc}$ value we obtained with low redshift data is consistent with the high redshift CMB observations($H_0 = 67.4\pm 0.5 \,Km/S/M_{pc}$) at one sigma~(\cite{2020A&A...641A...6P}).We also explored the late-time evolution of the universe corresponding to the best-fit model parameters.  Although, the formalism of this article is capable also for early scalar field dominated inflation, in the current work we focus on the study of the late time acceleration of the universe. For our analysis, we therefore, choose the scale factor between $a=0.1$ and $a=1.2$ which includes the current epoch $a_0 = 1$\footnote{The 
		further incorporation of  the early inflationary era will be followed in a future article.}.
	
	\vspace{5pt}
	The variation of equation of state parameter $\omega_{\phi}$ and deceleration parameter $q(a)$ for the best-fit parameters are shown in Fig.~\ref{wp_bestfit} and ~\ref{de_bestfit}. From the Fig.~\ref{wp_bestfit}, we can see that the present value($a=1$) of the equation of state parameter of the scalar field can go as low as $\omega_{\phi}=-0.9999$.  Thus at the present epoch,  the scalar field behaves just like a cosmological constant. This is because within two sigma limits the best-fit value of the power '$n$' is consistent with zero, see Table \ref{tab:best_fit_flat}, which gives a constant scalar field potential. Moreover, the transition between the deceleration to the accelerated phases of expansion occurs at $a\sim0.6$ which is also shown in Fig.~\ref{de_bestfit}. The Fig.~\ref{den_bestfit} shows the variation of energy density of the scalar field and the non-relativistic matter as a function of scale factor. We see that transition between matter 
	and dark-energy dominated Universe occurs around $a \sim 0.75$, i.e., the scalar field 
	starts to dominate at the very late stage of the evolution of the Universe and it drives the present accelerated expansion. From the Table~\ref{tab:best_fit_flat}, the present density parameter of the non-relativistic matter and dark energy corresponds to $\Omega_{m0}\sim 0.29$  and $\Omega_{\phi}\sim0.71$ respectively.
	
	\vspace{5pt}
	The method discussed in Sec.~\ref{sec:basic} is not limited to 
	the power law form of the potential that we have considered. It can also be applied to any other 
	forms of potentials which are not discussed here. This method helps in finding the cosmological dynamical 
	variables in an exact form without even knowing the evolution of scale factor. For instance, the Hubble paramter, $H(a)$ which we obtained analytically is directly observable. So the difficulty of solving coupled non-linear equations that one usually encounters while applying observational constraints in scalar field dark energy models can be alleviated. 
	

	\section*{Appendix A.}
	To derive Eqn.~\ref{V}  using Eqn.~\ref{phid2} we first obtain, 
	\begin{equation}
		\frac{d\dot \phi^2}{da} = 2\biggl[ \frac{dV}{da} \biggl(\frac{1+\omega_\phi}{1-\omega_\phi}\biggr)
		+V\bigg\{\frac{1}{1-\omega_\phi}+\frac{1+\omega_\phi}{\bigl(1-\omega_\phi\bigr)^2}\biggr\}\frac{d\omega_\phi}{da}\biggr] \, ,  
	\end{equation}
	which can be simplified as,
	
	\begin{eqnarray}
		\frac{d\dot \phi^2}{da} = \frac{2V}{\bigl(1-\omega_\phi\bigr)^2}\frac{d\omega_\phi}{da} + 
		2\frac{dV}{da} \biggl(\frac{1+\omega_\phi}{1-\omega_\phi}\biggr)\, .
		\label{dphi2da}
	\end{eqnarray}
	Using  Eqn.~\ref{dphi2da} in Eqn.~\ref{MasterEqn} and rearranging terms we obtain, 
	
	\begin{eqnarray}
		\begin{split}
			\frac{dV}{da}\biggl[1 + \frac{1+\omega_\phi}{1-\omega_\phi}\biggr] + \\&
			V\biggl[\frac{6}{a}\frac{\bigl(1+\omega_\phi\bigr)}{\bigl (1-\omega_\phi \bigr)}
			+ \frac{1}{\bigl( 1 - \omega_\phi\bigr)^2}\frac{d\omega_\phi}{da} \biggr]& = 0\, , 
		\end{split}	
	\end{eqnarray}
	
	which can be easily simplified into the form of  Eqn.~\ref{V}. 
	
		\section*{Acknowledgments}
	AJ acknowledges financial support from Ministry of Human Resource and Development, Government of India via Institute fellowship at IISER Bhopal.

	
	
\vspace{1cm}


\end{document}